# Work Design and Job Rotation in Software Engineering: Results from an Industrial Study

Ronnie E. S. Santos
Centro de Informática – Universidade Federal de Pernambuco
Brazil
ress@cin.ufpe.br

Maria Teresa Baldassarre
Dipartimento di Informatica – Università di Bari
Italy
mariateresa.baldassarre@uniba.it

Fabio Q. B. da Silva
Centro de Informática – Universidade Federal de Pernambuco
Brazil
fabio@cin.ufpe.br

Cleyton V. C. Magalhães
Centro de Informática – Universidade Federal de Pernambuco
Brazil
cvcm@cin.ufpe.br

Luiz Fernando Capretz
Electrical & Computer Eng. Dept.
Western University
Canada
lcapretz@uwo.ca

Jorge S. Correia-Neto
Distance Education and Technology Center – UFRPE
Brazil
jorgecorreianeto@gmail.com

*Abstract*—**Context: Job rotation is a managerial practice to be applied in the organizational environment to reduce job monotony, boredom, and exhaustion resulting from job simplification, specialization, and repetition. Previous studies have identified and discussed the use of project-to-project rotations in software practice, gathering empirical evidence from qualitative and field studies and pointing out set of work-related factors that can be positively or negatively affected by this practice. Goal: We aim to collect and discuss the use of job rotation in software organizations in order to identify the potential benefits and limitations of this practice supported by the statement of existing theories of work design. Method: Using a survey-based research design, we collected and analyzed quantitative data from software engineers about how software development work is designed and organized, as well as the potential effects of job rotations on this work design. We investigated 21 work design constructs, along with job burnout, role conflict, role ambiguity, and two constructs related to job rotation. Results: We identified one new benefit and six new limitations of job rotation, not observed in previous studies and added new discussions to the existing body of knowledge concerning the use of job rotation in software engineering practice. Conclusion: We believe that these results represent another important step towards the construction of a consistent and comprehensive body of evidence that can guide future research and also inform practice about the potential positive and negative effects of job rotation in software development companies.**

## I. INTRODUCTION

For decades, research on human resource management has investigated questions related to the work and discussed practices and approaches to improve individual performance. During this process, theories and discussions were developed, rising awareness about the way the work is performed. Viteles proposed one of the first theories about work design in the early 1950s [29]. Broadly, "work design" can be defined as different ways in which a given work or task can be designed, assigned to individuals and/or teams, and performed [18]. In general, this means the existence of processes and outcomes that encapsulate how the work is structured, organized, experienced, and legalized, including techniques to modify this structure, such as job simplification, job rotation, job enlargement, job enrichment, among others [1][29]. Job Rotation is one of the techniques applied to modify the way the work is structured, since it is defined as the systematic movement of employees from job to job, or project to project, within an organization, as a way to reduce the monotony, boredom and fatigue, resulting from job simplification, specialization, and repetition at work [4][30].

Recently, research regarding job rotation in the context of software engineering has demonstrated that the effects of this managerial technique are complex and potentially generate conflicting results to individuals working in software development. The results revealed the need to improve the understanding of work design in software engineering practice in order to advance on proposing or improving managerial techniques, such as job rotation in software companies. So far, the research has identified a set of work design factors, present in the software engineering context, which can be positively or negatively related to the rotation of individuals amongst software teams [23][24][25].

In the present study, our main goal is to extend the body of knowledge about the practice of job rotation in software companies by collecting data about work design and job rotation in the context of software engineering in order to answer the following research question:

*What are the correlations between job design characteristics and the use of project-to-project job rotation in software development industrial practice?*

To answer this question, we used a survey-based research to collect and analyze previously tested measures on work design characteristics, and then compared the findings with evidence from the previous research [23][24][25]. The development of such correlational study is important not only to increase list of known work elements affected by the practice of job rotation, but also to understand how these elements can interact. The correlation and the interaction among these factors were not explored in the previous studies. However, it is important to access it, in order to propose methods and models that could drive the plan and the configuration of job rotation in software companies. This configuration can be understood as the best way to perform a change in the composition of team working in a software project, considering to successfully moving individuals along with as less negative impact on their performance, motivation and satisfaction, as possible.

From this introduction, this paper is organized as follows. In Section 2, we present the conceptual background. In Section 3, we describe the research method, instruments and techniques applied to answer our research question. In section 4, we present the main findings, which are discussed in Section 5. Finally, in Section 6, we present our conclusions and directions for future research.

## II. Background

This section presents theoretical background about work design characteristics and job rotation in software engineering.

### A. Work Design and Job Characteristics

Work Design is defined as the process on how work should be conceived, assigned across organizational levels, and structured into tasks performed by individuals or teams [27] [10]. The interest to understand and improve work design in several types of organizations can be observed since the beginning of the 20th century, when Taylor [26] proposed a theory focused on simplification and specialization of work as an attempt to maximize workers' efficiency and productivity in organizations' mass production.

From that period, the work scenario in general has changed and different theories and approaches were needed to improve efficiency and productivity in workplace, depending on specific characteristics of each job [27]. In this sense, scientists and practitioners have been concerned about the need for instruments to assess work characteristics in organizations, such as the Job Diagnostic Survey [11], the Multi-method Job Design Questionnaire [2], and later, the integrative study of Morgenson and Humphrey to develop an instrument based on 107 terms of work characteristics found in an extensive literature review [17].

Morgenson and Humphrey's instrument is organized in 4 types of characteristics: Task, Knowledge, Social and Contextual, and has been applied in several types of organizations over the years [13]. Recently, we applied this questionnaire as a measure to investigate work design characteristics of software engineering, by exploring 21 work design constructs, together with job burnout, role conflict, role ambiguity and job rotation to validate the instrument in the context of software companies [5].

The results pointed out relevant characteristics of software engineering work and areas for further research. In particular, considering the different dynamics of the nature of work in software organizations, we believe that the study of work design and its relation with managerial techniques, such as job rotation, is relevant because the way the work is structured can affect several aspects of the organization and influence several work outcomes such as motivation, satisfaction or individual performance.

### B. Job Rotation in Software Engineering

Both types of job rotation defined by Woods [30] can be applied in the context of software companies, often, depending on the organizational needs [24]. For instance, in job-to-job (J2J) rotation, individuals are rotated between different jobs in the company, to perform activities with distinct natures and/or not directly related to the software development process, such as, a software developer that can be rotated to customer support department to increase knowledge redundancy at the organizational level [7]. In project-to-project (P2P) rotation, individuals are moved between projects of similar nature (e.g. two software development projects), often keeping the same technical role (e.g., a developer working with java in a project moved to work with C++ in other) or changing this role (e.g. a requirement analyst moved to a different project to work on acceptance testing) [23].

Recently, we have performed a set of qualitative studies related to the use of job rotation in the industrial practice of software engineering. So far, an industrial case study [23] and its extension [25], a systematic literature review [24], and a replication of the industrial case study [25] have led to a set of eight benefits, nine limitations, and two factors considered both benefits and limitations associated to the practice of job rotation in software engineering. The current list of benefits and limitations of job rotation identified so far is reproduced in TABLE I.

The present study is important to explore the correlations among job rotation and work design elements, since it was not explored in previous studies [23][24][25]. However, such correlations are important to understand the interaction between factors, turning feasible the proposition of theories and models that could guide practitioners in planning and configuring the best job rotation to each particular situation, depending on what they are seeking by applying such managerial approach.

TABLE I. BENEFITS AND LIMITATIONS OF JOB ROTATION IN SOFTWARE ENGINEERING (FROM SANTOS ET AL. [25])

| Factors | Effect of Job Rotation | Impact of Factor on Work | Benefit/ Limitation |
|---|---|---|---|
| *Organizational Factors* | | | |
| Communication | + | + | Benefit |
| Difficult to Plan | + | - | Limitation |
| Time Consuming | + | - | Limitation |
| *Team Factors* | | | |
| Team Flexibility | + | + | Benefit |
| Knowledge Exchange | + | + | Benefit |
| Knowledge Transfer | + | - | Limitation |
| Social Conflicts | + | - | Limitation |
| *Work Characteristics* | | | |
| *Task Characteristics* | | | |
| Task Variety | + | + | Benefit |
| Learning Opportunity | + | + | Benefit |
| Task Identity | - | + | Limitation |
| *Knowledge Characteristics* | | | |
| Skill Variety | + | + | Benefit |
| Specialization | - | +/- | Benefit/Limitation |
| *Social Characteristics* | | | |
| Feedback from others | - | + | Limitation |
| Social Support – Friendship Opportunities | +/- | + | Benefit/Limitation |
| *Outcomes* | | | |
| *Individual Outcomes* | | | |
| Motivation | + | + | Benefit |
| *Job Outcomes and Correlates* | | | |
| Job Monotony | - | - | Benefit |
| Cognitive Effort | + | - | Limitation |
| Workload | + | - | Limitation |
| Productivity | - | + | Limitation |

## III. METHOD

In this study, we applied a questionnaire to collect data from professional software engineers about their perceptions of the characteristics of their work and how these characteristics are related to job rotation. We investigated the correlations among work characteristics, its outcomes, and two dimensions of job rotation. In this process, we followed the definitions presented in the guidelines of Kitchenham and Pfleeger [21] and Linaker et al. [14] to perform cross-sectional surveys in software engineering, as well as our experience in conducting empirical studies in industry [31] [32].

Following the cited guidelines [14] [21], we designed our study following these methodological steps: Setting Objectives, Crafting Instrument, Data Collection, and Data Analysis. We describe the steps of this process below.

### A. Setting Objectives

In this study, we aimed to investigate the correlations among work design factors and the practice of job rotation seeking to understand the effects of this practice on the factors related to software engineers' work. We believe that a quantitative approach is important to identify correlations not revealed before in our previous studies, and also to compare the findings with the results obtained in the previous qualitative studies [23] [24] [25].

### B. Crafting Instrument

We built the questionnaire using tested measures and existing instruments to facilitate comparisons with related work and increase reliability, as discussed in the guidelines [21]. The only exception for the use of existing instruments was in one of the measures for job rotation, for which we developed three items to evaluate the degree of rotation experienced by professionals (Rotation Intensity), based on the script for interviews about this theme validated in our previous studies [23]. In other words, Rotation Intensity is a metric applied to verify how often software engineers have being rotated among different teams or projects in their company.

Thus, job characteristics were measured using the Work Design Questionnaire (WDQ) [17]. Further, to evaluate work outcomes, we used existing measures for satisfaction and job burnout. Satisfaction was measured using the Michigan Organizational Assessment Package [19]. To assess Job Burnout, we used the Maslach Burnout Inventory - General Survey [16]. This instrument measures three dimensions of job burnout: exhaustion (5 items), cynicism (5 items), and professional efficacy (6 items). We also measured role conflict and ambiguity, two correlation variables used by Hsieh and Chao [12], obtained from the Role Stress Assessment of Rizzo et al. [22].

Regarding job rotation, we adopted two sets of items to assess two dimensions of this practice. The first is related to the degree of rotation that the individual experiences in his/her job, which we named Rotation Intensity. For this dimension, we developed our own instrument that consists of three response items. The second is related to how easy or difficult it is to rotate individuals in a job or task. For this dimension, we used the Van de Ven and Ferry [28] Role Interchangeability measure, which was also used by Hsieh and Chao [12].

To build the questionnaire, we searched for validated versions of each instrument in Portuguese. As recommended in the guidelines [21] [14], we then performed a pilot study with 16 participants, among software engineering professionals and researchers. Results of the pilot test were used to clarify the wording of some sentences in Portuguese. Next, we tested the reliability and construct validity of all factors presented in the questionnaire on a sample of 77 professional software engineers and the results were published by da Silva et al. [5]. We used this validated instrument in this current study. All instruments are available in each of the cited studies [19] [16] [12] [22] [28].

TABLE II.  MEANS, STANDARD DEVIATIONS, AND RELIABILITY

| Construct | M | SD | α |
|---|---|---|---|
| *Task Characteristics* | | | |
| Work scheduling autonomy | 3,69 | 0,79 | 0.70 |
| Decision-making autonomy | 3,61 | 0,8 | 0.77 |
| Work methods autonomy | 3,66 | 0,78 | 0.76 |
| Task variety | 3,96 | 0,74 | 0.82 |
| Significance | 3,96 | 0,82 | 0.80 |
| Task identity | 3,82 | 0,75 | 0.72 |
| Feedback from job | 3,39 | 0,92 | 0.83 |
| *Knowledge characteristics* | | | |
| Job complexity | 3,61 | 0,74 | 0.62 |
| Information Processing | 4,18 | 0,59 | 0.66 |
| Problem solving | 3,85 | 0,66 | 0.55 |
| Skill variety | 4,11 | 0,71 | 0.84 |
| Specialization | 3,93 | 0,65 | 0.67 |
| *Social characteristics* | | | |
| Social support | 3,89 | 0,71 | 0.78 |
| Initiated interdependence | 3,51 | 0,85 | 0.71 |
| Received interdependence | 3,43 | 0,81 | 0.66 |
| Interaction outside organization | 3,11 | 1,14 | 0.86 |
| Feedback from others | 3,17 | 0,97 | 0.81 |
| Job Rotation | | | 0.66 |
| Rotation Use | 2,67 | 0.77 | 0.52 |
| Job Interchangeability | 2.87 | 0.76 | 0.52 |
| *Outcomes and Correlates* | | | |
| Job Burnout | 2,05 | 0.59 | 0.85 |
| Role Conflict | 2,47 | 0.76 | 0.78 |
| Role Ambiguity | 2,06 | 0.70 | 0.88 |
| Satisfaction | 4,19 | 0.79 | 0.78 |

*C. Data Analysis*

We performed data analysis similar to those used by Morgeson and Humphrey [17] and da Silva et al. [5]. We considered all scales to be interval, supported by the argument of Carifio and Perla [3] about Likert scales and Likert response items. We then redirected efforts to understand how the work design factors were co-related to the items that measured the two dimensions of job rotation.

To explain each factor identified as benefit or limitation in the context of software engineering, to allow comparison with previous studies, we applied techniques from meta-ethnography [6] to refine the meanings of these terms by enfolding the scientific literature on work design and organizational psychology and compared this to software development activities. Finally, to present these findings, we grouped the work-related factors affected by the practice of job rotation into two categories: a) Work Characteristics that include factors referring to the different ways in which a given work can be structured, assigned to individuals and/or teams, and performed, following the structure presented in the WDQ model; b) Outcomes, or factors related to the results of the work performed by an individual.

IV. RESULTS

We start this section presenting a brief description of the sample of individuals that participated in this study. We then present the statistical correlations among the two dimensions of job rotation and the factors analyzed in this study.

*A. Demographics*

The sample was composed by 126 professionals working in 40 different companies, in which 79% (99/126) were male and 21% (27/126) were female. Considering the role of individuals in software development, 50% (63/126) of our sample was composed of Program Developers, 29% (35/126) of Systems Analysts, 13% (17/126) of Software Testers, 5% (6/126) of Project Managers, and 4% (5/126) of UX/UI Designer. Regarding the experience in software industry, 26% of the sample (33/126) was working in software development for less than 5 years, 31% of individuals (39/126) had been working in this field for a period between 5 and 10 years, and 43% of participants (54/126) had more than 10 years of experience in software development. TABLE III. summarizes this information..

TABLE III.  SUMMARY OF PARTICIPANTS ON THIS STUDY

| | | | Age | | Job Experience (years) | | Sex |
|---|---|---|---|---|---|---|---|
| Role | Total | % (N=126) | M | SD | M | SD | %men |
| Analyst | 35 | 28% | 36,5 | 9,6 | 13,1 | 13,1 | 21% |
| Manager | 6 | 5% | 45,2 | 6,7 | 22,7 | 22,7 | 1% |
| Tester | 17 | 13% | 35,9 | 3,9 | 8,4 | 8,4 | 6% |
| Developer | 63 | 50% | 31,5 | 5,4 | 8,9 | 8,9 | 47% |
| Designer | 5 | 4% | 37,8 | 4,9 | 8,6 | 8,6 | 3% |
| Total | 126 | 100% | | | | | 79% |

*B. Correlations between Job Rotation and Work Characteristics*

The main results were obtained using the Spearmans's ρ correlation. We used the same categories used in the WDQ model constructed by Morgeson and Humphrey [17] [18] to organize and present these findings. At the highest level, benefits and limitations were split into Work Characteristics and Outcomes. The abbreviations RI and JI refer to the two dimensions of job rotation: Rotation Intensity and Job Interchangeability, respectively. To define the impact of each factor on the work of software engineers, we followed the descriptions in the literature and in the instruments applied. Work design theories contend that work characteristics are all beneficial to the work; therefore, we considered the impact of all characteristics as positive when determining the benefits or limitations of job rotation. On the other hand, Role Conflict and Job Burnout are associated with negative impact on the work, and Satisfaction as positive.

Rotation Intensity showed a significant negative correlation with Task Identity, Feedback from the Job, Job Complexity and Satisfaction, together with a positive correlation with Job Burnout. On the other hand, Job Interchangeability presented significant negative correlations with Information Processing, Initiated Interdependence and Role Conflict, as summarized in TABLE IV. Due to template restrictions, correlations obtained in this study are not presented in this paper.

The negative correlation between Rotation Intensity and Task Identity confirms and reinforces the results of previous studies [23] [25]. As observed before, Task Identity reflects the degree to which a job involves a whole piece of work, that is, a well-defined work. Therefore, the negative correlation emphasizes a reduction on the perception that software engineers have in understanding their work as a process with beginning, middle, and end, which can directly impact their motivation at work [9]. Job Rotation has a negative impact on this factor, and practitioners must be aware of that, to mitigate potential negative impacts on motivation.

Rotation Intensity is negatively correlated with Feedback from Job. This factor reflects the degree to which the work itself can provide information about the effectiveness of professionals while performing a specific task [17]. Consequently, a negative correlation between a dimension of Job Rotation and Feedback from Job means that software engineers seem to perceive less feedback resulting directly from their job activities in the context of rotations at work. This is important because previous studies have demonstrated the importance of feedback for the satisfaction of software engineers [8] [9].

Both dimensions of job rotation (RI and JI) demonstrated a significant negative correlation with factors related to knowledge characteristics of work. Rotation Intensity is negatively correlated with Job Complexity, while Job Interchangeability is negatively related with Information Processing. Following the work design theory, Job Complexity refers to complex tasks that require high-level skills of professionals at work, while Information Processing reflects the degree to which a job requires active information processing to be performed. Both factors are related to creative process and result in positive motivational outcomes [17]. Therefore, considering previous findings [25], we believe that this negative correlation is related to the amount of variety of tasks and skills involved in the rotation process. When a rotation involves low levels of variety related to role, type of software project or business domain, the individual experiences less opportunities to work with complex and creative tasks, thus, experiencing less benefits resulting from the rotation, concerning knowledge factors. In fact, our previous studies discussed that practitioners must consider the amount of variety involved when planning a rotation due to its potential effects on the knowledge characteristics of work [23] [25].

Considering social characteristics of work, the Job Interchangeability dimension of job rotation showed significant negative correlation with Initiated Interdependence. In the work design theories, interdependence reflects the degree to which the job depends on others and others depend on it to complete the work. The Initiated Interdependence can be defined as the extent to which work flows from one job to other jobs [17]. For instance, the existing connection among the end of the activities of software requirements and the beginning of the work of developers in the project. Our results demonstrated that the practice of job rotation could negatively influence this aspect of work, and therefore, represent a limitation of this practice because professionals can lose the perception of work flows from their job to the job of other professionals participating in the process, due to frequent changes on the team.

Regarding work outcomes, we observed positive correlation between Rotation Intensity and Job Burnout, and also negative correlation between Rotation Intensity and Satisfaction. These results support the early claims related to indirect effect of job rotation in the satisfaction of software engineers, considering that this practice influences work-factors that have direct impact on Job Satisfaction, such as feedback and performance [23][9]. Further, this result confirms the discussions of Hsieh and Chao [12] that argued that the effect of job rotation is consistent with dimensions of Job Burnout (exhaustion, cynicism, and professional efficacy). However, we need to further explore this correlation, considering that many other studies demonstrated the opposite relation among job rotation and job burnout [5] [30]. Finally, we found a negative correlation of Job Interchangeability with Role Conflict, which is defined as the results from two or more sets of incompatible demands involving work-related issues [22]. This result points out a benefit of job rotation to software companies, since the evidence indicate that this practice can reduce this type of conflict that is prejudicial for work.

In summary, all the correlations identified in the full sample were here described and compared with results of previous studies, resulting in consistent and comprehensive list of extra information to the body of evidence about the effects of job rotation in the software engineering practice.

TABLE IV. NEW IDENTFIED BENEFITS AND LIMITATION OF JOB ROTATION

| Factors | Correlation with factor | Impact of factor on the work | Benefit /Limitation |
|---|---|---|---|
| *Work Characteristics* | | | |
| *Task Characteristics* | | | |
| **Task Identity** | -(RI)* | + | **Limitation** |
| **Feedback from Job** | -(RI)** | + | **Limitation** |
| *Knowledge Characteristics* | | | |
| **Job Complexity** | -(RI)* | + | **Limitation** |
| Information Process | -(JI)** | + | Limitation |
| *Social Characteristics* | | | |
| Initiated interdependence | -(JI)** | + | Limitation |
| *Outcomes* | | | |
| Role Conflict | -(JI)* | - | Benefit |
| Job Burnout | +(RI)* | - | Limitation |
| Satisfaction | -(RI)* | + | Limitation |

## V. DISCUSSIONS

In this section, we discuss the findings of our research along with the implications of the results for research and practice.

### A. Comparing Findings and Implications

In this study, we extended the existing list of known factors present in the software development context that can be potentially affected by the use of job rotation in practice. One of the main contributions to academic research that can be highlighted is the importance of applying mixed-methods to conduct research in Software Engineering. We observed that the results obtained from previous qualitative studies were probably converging because of the subjective perceptions of participants concerning the phenomenon under investigation. The use of a different approach was effective in getting a complementary perspective from the professionals.

This study demonstrated significant differences between the results obtained through a quantitative approach and the evidence gathered in previous qualitative studies. Both studies performed previously (Case Study and SLR) and this survey research presented only one common factor affected by the practice of job rotation in software companies: Task Identity.

For practitioners, the main result of this study is the identification of eight correlations among job rotation, work-related factors and work outcomes, adding one new benefit and six new limitations on the set of previous known factors affected by job rotation and presented in Table I. We believe that the different perspectives that are evaluated in the two types of research methods can explain these results. In the qualitative research, we looked for the subjective perceptions of the software engineers based on their feelings and emotions related to their experience with job rotation. On the other hand, the quantitative research tried to assess the objective characteristics of the work, not the individual feelings or attitudes towards these characteristics.

Therefore, it is plausible that the emotional or attitudinal responses to the work characteristics are different from the objective assessment of the same characteristic. For instance, the software engineer may have a very high positive attitude towards Task Variety, one important factor identified in all three previous studies, due to his/her desire to perform different tasks on the job, supporting Task Variety being spontaneously associated with job rotation in the case studies. However, higher levels of job rotation may not correlate with higher levels of Task Variety because the software engineer may objectively assess his/her work as having high levels of variety regardless of the level of rotation experienced in the workplace.

We understand that this extensive list of factors has a certain level of difficulty to be managed, considering practical applications. However, complex phenomena "in the wild" usually entail this kind of complexity. The goal of our research is the construction of a consistent and comprehensive body of evidence that can on one hand guide future research and, on the other hand, be sufficiently complete to inform practice about the potential positive and negative effects of job rotations.

Moreover, this study adds important new evidence that can be applied in the process of plan a job rotation, depending on what outcome practitioners are targeting. For instance, if a rotation is planned to be performing targeting the increase on the motivation of a professional, the process might consider the correlations among the work elements that could influence one's motivation, for instance, variety of tasks or learning opportunities. However, since several factors interact together, practitioners need to be aware of the negative impacts of such rotation, and plan the ideal configuration depending on each specific case, individual, and required outcome, such as performance, satisfaction, or individual needs.

Further investigation on the interacting effects of these factors is indeed needed towards the definition of a theory that can explain this phenomenon in software engineering practice and also to support the construction of guidelines or other managerial tools to support practitioners. However, the results presented in this article are a step forward in this direction, particularity because previous studies had pointed out that a rotation can be configured considering basically two triggers, namely, Project Needs and Individual Requests [23]. However, the correlations found in this study demonstrated that to plan a rotation in software engineering software project managers need to consider: a) Rotation Intensity, which means that practitioners need to observe the software project phase and the frequency of rotation that one are experiencing, for example, recently rotated or never rotated, and b) Job Interchangeability, which reflects the level of variety among tasks before and after the rotation and the amount of information that needs to be processed for one to fit in the work of another.

Finally, this study reinforces that the practice of job rotation is intimal related to personal human and emotional aspects of work, being related to elements such as, motivation, satisfaction, burnout, conflicts and so on. Therefore, it confirms the importance of creating and evaluating a theory and guidelines to inform and guide practice.

### B. Threats to Validity

Considering construct validity, we applied measures and instruments that had been successfully tested and applied in previous studies, including in the context of software engineering [6]. In particular, the data collected in this study demonstrated good internal consistency.

We did not use a random sampling strategy from a well-defined population. Therefore, we cannot claim statistical generalization to any large population, which was not our goal from the beginning. Nevertheless, the model obtained from the results can be used for analytical generalization and hypothesis building that can be used to guide future research. Our sample is not uniform regarding demographics such as gender, age, etc. If the perceptions of work characteristics differ between different categories of individuals regarding such variables, our results could be biased. However, previous theories of work design do not show significant differences in perception regarding these variables.

One important threat to external validity is that our sample is from professionals working in Brazilian software companies. Different cultural practices and issues such as the legal framework that regulates job relationships might have influenced the results. However, it is important to highlight that our sample included individuals working in branches and offices of global software companies, therefore many individuals had a perception based on the framework of other countries. In addition, to further strengthen the results of this study, we are currently planning a replicated multi-country study, involving software organizations from Brazil, Canada, and Italy.

## VI. Conclusions

We identified evidence about the benefits and limitations of job rotation in software engineering practice using an approach and method different from those previously applied in this context. In this study, one new benefit and six new limitations were added to the existing list of known factors affected by this practice. They should be considered by practitioners when planning or applying rotation of individuals among software projects and teams. We found no contradictory results with those previously published in our research [23] [24] [25], which contributes to the strength of the evidence found so far. This emphasizes the importance of conducting mix-method research to investigate a phenomenon that involves multiple variables.

Our findings show that project-to-project rotations are related to important benefits regarding knowledge characteristics and outcomes of the work in software organizations, such as the potential reduction of role conflict. We also identify limitations that can be observed at the work level and outcomes that raised some concern about how this practice is affecting the satisfaction of software engineers. We discussed above that these factors have interacting effects that create a challenge for the effective use of job rotations in the software engineering practice.

Our main goal is to inform practitioners that they must be aware of these challenges and carefully devise a plan for rotations considering a broad set of factors. Some of these challenges must be addressed in our future research, involving not only job rotation in particular, but also broad aspects of work design in software engineering. In particular, we understand that the process for the construction of a theory and guidelines regarding this practice will involve the interacting effects of multiple factors. Therefore, this process shall be addressed by using multiple research methods as well as different qualitative and quantitative data. The development of longitudinal studies is important and still required to check the interplay between short-term negative effects and long-term benefits of job rotation.

### Acknowledgment

Fabio Q. B. da Silva holds a research grant from CNPq #314523/2009-0. Cleyton V. C. Magalhães and Ronnie E. S. Santos are PhD students and receive a scholarship from CNPq. This work is the result of a MoU between University of Bari and Federal University of Pernambuco. Last but not least, we are very grateful to all participants of this study.